\documentclass[aps,10pt,pra,twocolumn,amsmath,amssymb,superscriptaddress,groupedaddress]{revtex4-1}

\usepackage[english]{babel} 
\usepackage[utf8]{inputenc}
\usepackage[T1]{fontenc}
\usepackage{graphicx}
\usepackage{epsfig}
\usepackage{graphicx}
\usepackage{subfigure}

\usepackage{hyperref}
\hypersetup{
	colorlinks=true,
	linkcolor=blue,			
	citecolor=green,		
	filecolor=magenta,		
	urlcolor=magenta		
}

\newcommand{\ket}[1]{\left\vert #1 \right\rangle}
\newcommand{\bra}[1]{\left\langle #1 \right\vert}
\newcommand{\bracket}[2]{\left \langle #1 \vert #2 \right\rangle}
\newcommand{\abs}[1]{\left\vert #1 \right\vert}
\newcommand{\erw}[1]{\left\langle #1 \right\rangle}
\newcommand{\komm}[2]{[ #1 , #2 ]}
\newcommand{\akomm}[2]{\lbrace #1, #2 \rbrace}
\newcommand{\D}[1]{\mathcal{D}[#1]}
\newcommand{\super}[1]{\mathcal{#1}}
\newcommand{\hc}{\mathrm{H.c.}}
\renewcommand{\d}{\mathrm{d}}
\newcommand{\Exp}{\operatorname{E}}
\newcommand{\comma}{~,}
\newcommand{\fullstop}{~.}
\newcommand{\Tr}{\mathrm{Tr}}
\renewcommand{\Re}{\operatorname{Re}}
\renewcommand{\Im}{\operatorname{Im}}

\begin{document}

\selectlanguage{english}

\title{Heralded dissipative preparation of nonclassical states in a Kerr oscillator}

\author{Martin Koppenh\"ofer}
\affiliation{Department of Physics, University of Basel, Klingelbergstrasse 82, CH-4056 Basel, Switzerland}

\author{Christoph Bruder}
\affiliation{Department of Physics, University of Basel, Klingelbergstrasse 82, CH-4056 Basel, Switzerland}

\author{Niels L\"orch}
\affiliation{Department of Physics, University of Basel, Klingelbergstrasse 82, CH-4056 Basel, Switzerland}

\date{\today}

\begin{abstract}
We present a heralded state preparation scheme for driven nonlinear open quantum systems. 
The protocol is based on a continuous photon counting measurement of the system's decay channel.
When no photons are detected for a period of time, the system has relaxed to a measurement-induced pseudo-steady state.
We illustrate the protocol by the creation of states with a negative Wigner function in a Kerr oscillator, a system whose unconditional steady state is strictly positive.
\end{abstract}

\maketitle

\def\ValueFigIGammaRel{0.39}
\def\ValueFigIGammaJump{0.39}
\def\ValueFigINtraj{500}
\def\ValueFigIDelta{1.5}
\def\ValueFigIK{2.2}
\def\ValueFigIkappa{1.0}
\def\ValueFigIalphaI{0.82}
\def\ValueFigIalphaII{0}
\def\ValueFigIdelta{1.5}
\def\ValueFigIP{1.5}
\def\ValueFigIeta{1}
\def\ValueFigInth{0}
\def\ValueFigIxi{0}
\def\ValueFigIminWaitingTime{5}

\def\ValueFigIIdelta{1.5}
\def\ValueFigIIKopt{2.2}
\def\ValueFigIIkappa{1.0}
\def\ValueFigIIP{1.5}
\def\ValueFigIIeta{1}
\def\ValueFigIInth{0}
\def\ValueFigIIxi{0}
\def\ValueFigIIxiOptAbs{0.91}
\def\ValueFigIIxiOptAngle{1.78}

\def\ValueFigIIINtraj{500}
\def\ValueFigIIIDelta{0}
\def\ValueFigIIIK{10}
\def\ValueFigIIIkappa{1}
\def\ValueFigIIIaI{0}
\def\ValueFigIIIaII{5.3}
\def\ValueFigIIInth{0}
\def\ValueFigIIIeta{1}
\def\ValueFigIIIxi{0}
\def\ValueFigIIIGammaJump{0.42}
\def\ValueFigIIIGammaAsy{0.42}
\def\ValueFigIIIGammaRel{0.78}

\def\ValueFigIVSCDelta{1.5}
\def\ValueFigIVSCK{2.2}
\def\ValueFigIVSCkappa{1.0}
\def\ValueFigIVSCaI{0.82}
\def\ValueFigIVSCaII{0}
\def\ValueFigIVSCxiabs{0.9}
\def\ValueFigIVSCxiarg{1.8}
\def\ValueFigIVSCdelta{1.5}
\def\ValueFigIVSCP{1.5}
\def\ValueFigIVPADelta{0}
\def\ValueFigIVPAK{10}
\def\ValueFigIVPAkappa{1.0}
\def\ValueFigIVPAaI{0}
\def\ValueFigIVPAaII{5.3}
\def\ValueFigIVPAxi{0}
\def\ValueFigIVSCnthI{0}
\def\ValueFigIVSCetaI{1}
\def\ValueFigIVSCnthII{0.5}
\def\ValueFigIVSCetaII{1}
\def\ValueFigIVSCnthIII{0}
\def\ValueFigIVSCetaIII{0.25}
\def\ValueFigIVPAnthI{0}
\def\ValueFigIVPAetaI{1}
\def\ValueFigIVPAnthII{0.1}
\def\ValueFigIVPAetaII{1}
\def\ValueFigIVPAnthIII{0}
\def\ValueFigIVPAetaIII{0.5}

\section{Introduction}
Nonlinearity is a crucial prerequisite for quantum algorithms to outperform their classical counterparts in quantum information processing because it gives rise to states or operations that cannot be efficiently described in a classical framework \cite{NielsenChuang}. 
An important property to evaluate the usefulness of a quantum state in this context is the occurrence of negative values in its phase-space quasiprobability distribution \cite{Veitch-NJP.14.113011,Stahlke-PhysRevA.90.022302,Saleh-PhysRevX.6.021039}. 

However, such nonclassical states are challenging to prepare and stabilize because of unavoidable decoherence due to interaction with an unmonitored environment. 
For example, the perhaps simplest nonlinear quantum system, a driven and damped quantum oscillator with a Kerr nonlinearity, has a steady-state  Wigner function that is strictly positive \cite{Kheruntsyan-OptCommun.139.157,Kheruntsyan-JourOptB.2.225,Bartolo-PhysRevA.94.033841}. 

Here, we circumvent this restriction and quantify the potential of such a system to stabilize nonclassical states with negative Wigner density.  
We consider setups where a detector continuously monitors the emitted photons.
Such information leaking out of the system has already been useful in the context of 
entanglement generation 
\cite{PhysRevA.59.1025,PhysRevA.59.2468,PhysRevLett.93.120408,PhysRevLett.98.190501,PhysRevLett.105.210502},
teleportation \cite{PhysRevLett.83.5158}, 
cooling \cite{PhysRevLett.80.688,PhysRevA.60.2700,PhysRevB.68.235328,Nature.524.325},
and nonclassical optomechanical limit cycles \cite{PhysRevA.97.063812}, 
since the continuous observation modifies the system's dynamics.
In general, the states of the system during a continuous monitoring can have negative Wigner densities, but they fluctuate stochastically and feedback protocols are necessary to stabilize a particular state \cite{Minganti-SciRep.6.26987}.
In this work, we demonstrate that quantum trajectories can continuously relax to deterministic states whose presence is revealed by the detection signal. 
This mechanism opens a new alternative path in heralded quantum state preparation and allows one to stabilize certain nonclassical states without feedback, including Schr\"odinger kitten states.

In contrast to most heralded state preparation protocols relying on a photon detection event that heralds the \emph{projection} to a (potentially maintained) target state \cite{Clausen-JOptB.1.332,Lund-PhysRevA.70.020101,Bimbard-NatPhot.4.243,Hong-science.358.203,Lance-PhysRevA.73.041801,Galland-PhysRevLett.112.143602,Takeda-PhysRevA.87.043803,Baranger-PhysRevLett.122.140502}, we explore  the opposite approach and use the photon-counting measurement to identify a time evolution which continuously \emph{relaxes} the system into the target state, similar to Ref.~\onlinecite{Soergel}. 
Because the system will stay in this state conditioned on no further photon detection events, we will refer to it as a pseudo-steady state, to distinguish our mechanism from dissipative steady-state stabilization \cite{Poyatos-PhysRevLett.77.4728,Krauter-PhysRevLett.107.080503,Mamaev-quantum.2.58,Brunelli-PhysRevA.98.063801}. 

On one hand, our results shed light on the actual dynamics of an open quantum system when the information leaking out to the environment is not discarded. 
On the other hand, they can be seen as a practical protocol for heralded state preparation in open quantum systems that is feasible with current technology.

\section{System}
\begin{figure}
	\centering
	\includegraphics[width=.48\textwidth]{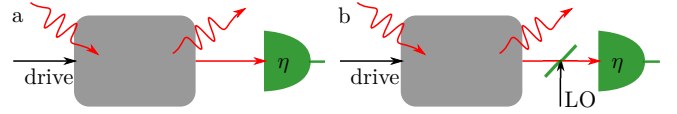}
	\caption{
		(a) A driven nonlinear open quantum system (gray box) is monitored by a photon-counting measurement of detection efficiency $\eta$.
		The detection signal provides a herald for the creation of a pseudo-steady state in the system. 
		(b) In a homodyne detection setup, a local oscillator (LO) signal is added before the detection, which allows one to modify the pseudo-steady state. 
	}
	\label{fig:System}
\end{figure}
We consider an open quantum system exchanging photons with a finite-temperature environment. 
Its quantum master equation is ($\hbar = 1$)
\begin{align}
	\frac{\d}{\d t} \hat{\rho} =  \super{L}_0 \hat{\rho} + \kappa (n_\mathrm{th} + 1) \D{\hat{a}} \hat{\rho} + \kappa n_\mathrm{th} \D{\hat{a}^\dagger} \hat{\rho} \comma
	\label{eqn:QME}
\end{align}
where $\hat{a}$ is the photon annihilation operator, $\kappa$ denotes the decay rate, $n_\mathrm{th}$ is the thermal photon number, and $\D{\hat{O}} \hat{\rho} = \hat{O} \rho \hat{O}^\dagger - \akomm{\hat{O}^\dagger \hat{O}}{\hat{\rho}}/2$ is a Lindblad dissipator.
In general, $\super{L}_0$ can be any completely positive and trace-preserving linear superoperator such that Eq.~\eqref{eqn:QME} has a steady-state solution $\hat{\rho}_\mathrm{ss}$.
For now, we choose $\super{L}_0 \hat{\rho} = - i \komm{\hat{H}_0}{\hat{\rho}}$, where 
\begin{align}
	\hat{H}_0 = - \Delta \hat{a}^\dagger \hat{a} + K \hat{a}^\dagger \hat{a}^\dagger \hat{a} \hat{a} + \left( \alpha_1 \hat{a}^\dagger + \alpha_2 \hat{a}^\dagger \hat{a}^\dagger + \hc \right)
	\label{eqn:Hamiltonian}
\end{align}
describes an anharmonic oscillator with a Kerr nonlinearity of strength $K$ that is subjected to semiclassical and parametric drives of strength $\alpha_1$ and $\alpha_2$, respectively.
We work in a frame rotating at the semiclassical drive frequency $\omega_\mathrm{drive}$, and $\Delta = \omega_\mathrm{drive} - \omega_0$ is the detuning with respect to the natural frequency $\omega_0$. 
The photon emission of the system is constantly monitored by a photon detector, as shown in Fig.~\ref{fig:System}(a). 
To illustrate the basic principle of the protocol, we first focus on the case of a zero-temperature environment, $n_\mathrm{th} = 0$, and unit detection efficiency of the photon-counting measurement, $\eta = 1$. 
The effect of finite temperature and imperfect detection is discussed in section~\ref{sec:FINITE}, and a detailed study including a more general form of $\super{L}_0$ is given in Appendix~\ref{sec:app:PseudoStationaryStateAndRelaxationRate}.

To model the photon-counting measurement, Eq.~\eqref{eqn:QME} is rewritten to a stochastic Schr\"odinger equation \cite{WisemanMilburn},
\begin{align}
	\d \ket{\psi} 
	&= \super{H} \ket{\psi} \d t 
	+ \left( \frac{\hat{a} \ket{\psi}}{\sqrt{\bra{\psi} \hat{a}^\dagger \hat{a} \ket{\psi}}} - \ket{\psi} \right) \d N \fullstop
	\label{eqn:SSE}
\end{align}
The term in brackets describes sudden quantum jumps of the state vector $\ket{\psi}$ due to photon detection events. 
The Poissonian stochastic increment $\d N$ is unity if the photon detector clicks and zero otherwise. 
It has an ensemble-averaged expectation value $\Exp(\d N) = 2 \bra{\psi} \hat{M} \ket{\psi} \d t$, where we have introduced the abbreviation $\hat{M} = \kappa \hat{a}^\dagger \hat{a}/2$. 
The continuous time evolution of $\ket{\psi}$ in the absence of photon detection events is captured by the nonlinear operator 
\begin{align}
	\super{H} \ket{\psi} = [ -i (\hat{H}_0 - i \hat{M}) + \bra{\psi} \hat{M} \ket{\psi} ] \ket{\psi} \fullstop
	\label{eqn:HNL}
\end{align}
The non-Hermitian correction $- i \hat{M}$ to the Hamiltonian $\hat{H}_0$ introduces relaxation and a decay of the norm of $\ket{\psi}$. 
To preserve the norm, we include the nonlinear term $\bra{\psi} \hat{M} \ket{\psi} \ket{\psi}$ in $\super{H}$. 
By construction, an ensemble average over many solutions of Eq.~\eqref{eqn:SSE} for different realizations of the stochastic jump process, so-called quantum trajectories, recovers the solution of Eq.~\eqref{eqn:QME} \cite{WisemanMilburn}. 
Note that the time evolution with $\mathcal{H}$ corresponds to the rare-event limit $s \to \infty$ in a generalized master equation \cite{PhysRevLett.104.160601,PhysRevLett.100.150601,PhysRevLett.90.238305}.

\section{Protocol}
\label{sec:PRT}
\begin{figure}
	\centering
	\includegraphics[width=.48\textwidth]{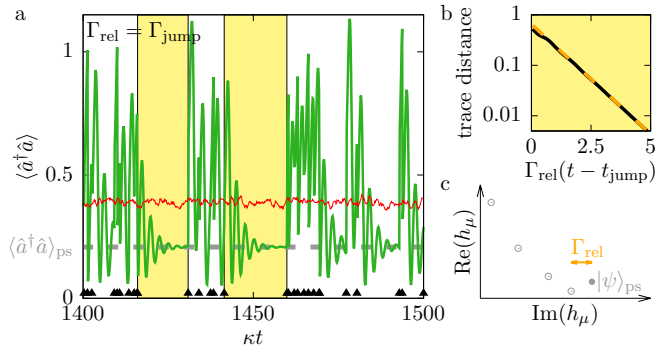}
	\caption{
		(a) Steady-state dynamics of the photon-number $\erw{\hat{a}^\dagger \hat{a}}$ of a Kerr oscillator subjected to a semiclassical drive. 		
		An average over $\ValueFigINtraj$ quantum trajectories reproduces the constant steady-state result $\erw{\hat{a}^\dagger \hat{a}}_\mathrm{ss}$ (thin solid red line), which determines the average photon detection rate $\Gamma_\mathrm{jump}$. 
		Along a single quantum trajectory, $\erw{\hat{a}^\dagger \hat{a}}$ (solid green line) evolves by stochastic quantum jumps at a rate $\Gamma_\mathrm{jump}$ (jump times indicated by black triangles) interchanged with a relaxation toward a pseudo-steady state $\ket{\psi}_\mathrm{ps}$ at a rate $\Gamma_\mathrm{rel}$. 
		The corresponding photon number $\erw{\hat{a}^\dagger \hat{a}}_\mathrm{ps}$ is marked by the thick dashed gray line. 
		In the intervals highlighted in yellow (light gray), the waiting time between two adjacent quantum jumps is longer than $\ValueFigIminWaitingTime$ times the relaxation time.
		(b) The trace distance between the state $\ket{\psi(t)}$ and $\ket{\psi}_\mathrm{ps}$ (solid black line) decays exponentially after a quantum jump event.
		The decay rate is $\Gamma_\mathrm{rel}$ (dashed orange line).  
		(c) Spectrum of the non-Hermitian Hamiltonian that defines the relaxation dynamics. 
		The relaxation rate $\Gamma_\mathrm{rel}$ is the imaginary part of the smallest gap between the stable eigenstate $\ket{\psi}_\mathrm{ps}$ (solid circle) and the unstable eigenstates (open circles). 
		Parameters: $\Delta/\kappa = \ValueFigIdelta$, $K/\kappa = \ValueFigIK$, $\abs{\alpha_1}^2 K/\kappa^3 = \ValueFigIP$, $\alpha_2/\kappa = \ValueFigIalphaII$, and $\xi = \ValueFigIxi$. 
	}
	\label{fig:Protocol}
\end{figure}

The stochastic Schr\"odinger equation~\eqref{eqn:SSE} describes a continuous time evolution of the state $\ket{\psi}$ that is interrupted by discontinuous quantum jump events.
This will lead to an interplay of two timescales:
After initial transient dynamics, the quantum trajectories fluctuate on average around the steady state $\hat{\rho}_\mathrm{ss}$ of Eq.~\eqref{eqn:QME}, as shown in Fig.~\ref{fig:Protocol}(a). Quantum jumps happen at a rate $\Gamma_\mathrm{jump} = 2 \Tr ( \hat{M} \hat{\rho}_\mathrm{ss})$. 
Between two adjacent quantum jumps, the state $\ket{\psi}$ evolves continuously according to the operator $\super{H}$, which has a steady-state solution fulfilling $\super{H} \ket{\psi}_\mathrm{ps} = 0$ and an associated relaxation rate $\Gamma_\mathrm{rel}$. 
In the following, we will call $\ket{\psi}_\mathrm{ps}$ the \emph{pseudo-steady state} of the stochastic Schr\"odinger equation~\eqref{eqn:SSE} because it is a steady state conditioned on the absence of photon detection events. 
In the regime $\Gamma_\mathrm{rel} \gtrsim \Gamma_\mathrm{jump}$, the waiting time between two adjacent quantum jumps can be much larger than the relaxation time and $\ket{\psi}$ relaxes exponentially to $\ket{\psi}_\mathrm{ps}$, as shown in Fig.~\ref{fig:Protocol}(b).
Hence, a photon detection event followed by no further click of the detector for several relaxation times $1/\Gamma_\mathrm{rel}$ heralds the preparation of the state $\ket{\psi}_\mathrm{ps}$ and the waiting time since the last detection event determines the state preparation fidelity.

\section{Pseudo-steady state and relaxation rate} 
\label{sec:PSS}
We now derive explicit expressions for the pseudo-steady state and the relaxation rate. 
We assume that the non-Hermitian operator $\hat{H}_0 - i \hat{M}$ has a set of left and right eigenvectors that can be normalized to form a complete orthonormal basis. 
The complex spectrum of $\hat{H}_0 - i \hat{M}$ is denoted by $\{ h_\mu \}$, \emph{i.e.}, 
\begin{align}
	(\hat{H}_0 - i \hat{M}) \ket{\psi_\mu} = h_\mu \ket{\psi_\mu} \fullstop
	\label{eqn:App:B:BasisKets}
\end{align}
A pseudo-steady state of Eq.~\eqref{eqn:SSE} is a normalized state vector $\ket{\psi}$ that satisfies $-i E_\psi \ket{\psi} = \super{H} \ket{\psi}$, where $E_\psi$ is real.
Such a solution can exist because the nonlinear term in Eq.~\eqref{eqn:HNL} compensates the decay of the norm induced by $-i \hat{M}$. 
To find the pseudo-steady state solution $\ket{\psi}$, we decompose $\ket{\psi} = \sum_\mu c_\mu \ket{\psi_\mu}$ with respect to the basis of eigenvectors $\ket{\psi_\mu}$ and obtain the following conditions for the expansion coefficients $c_\mu$: 
\begin{align}
	\forall \mu : \quad c_\mu \left[ - i (E_\psi - h_\mu) - \sum_{\beta,\gamma} c_\beta^* c_\gamma \bra{\psi_\beta} \hat{M} \ket{\psi_\gamma} \right] = 0 \fullstop
	\label{eqn:App:B:Coefficients}
\end{align}

For a non-degenerate eigenvalue $h_\nu$, Eq.~\eqref{eqn:App:B:Coefficients} implies that all expansion coefficients 
are zero except for the coefficient $c_\nu = 1/\sqrt{\bracket{\psi_\nu}{\psi_\nu}}$ of the corresponding eigenstate $\ket{\psi_\nu}$.
Thus, each normalized eigenstate $\ket{\psi_\nu}$ to a non-degenerate eigenvalue $h_\nu$ is a pseudo-steady-state solution with real energy $E_{\psi_\nu} = \bra{\psi_\nu} \hat{H}_0 \ket{\psi_\nu}$. 
For a degenerate eigenvalue $h = h_{\nu_1} = \dots = h_{\nu_N}$, any normalized superposition $\ket{\psi} = \sum_{i=1}^N c_{\nu_i} \ket{\psi_{\nu_i}}$ of the eigenstates belonging to this degenerate subspace is a pseudo-steady state with $E_\psi = \bra{\psi} \hat{H}_0 \ket{\psi}$. 

Since $\super{H}$ is a nonlinear operator, some of the pseudo-steady states $\super{H} \ket{\psi} = -i E_\psi \ket{\psi}$ may be unstable. 
To analyze the stability of a pseudo-steady state $\ket{\psi}$ with associated eigenvalue $h$, we make the ansatz
\begin{align}
	\ket{\chi} = e^{-i E_\psi t} (\ket{\psi} + \varepsilon \ket{\sigma})[ 1 - \varepsilon \Re( \bracket{\psi}{\sigma} ) ] \comma
\end{align}
where $\varepsilon \ll 1$ is a small parameter and $\ket{\sigma}$ is a state orthogonal to $\ket{\psi}$. 
Note that $\ket{\chi}$ is normalized to leading order in $\varepsilon$. 
We now expand $\d \ket{\chi} = \super{H} \ket{\chi} \d t$ in powers of $\varepsilon$ and decompose $\ket{\sigma} = \sum_\mu c_\mu \ket{\psi_\mu}$ with respect to the basis of eigenstates $\ket{\psi_\mu}$ of $\hat{H}_0 - i \hat{M}$, which yields
\begin{align}
	\sum_\mu \dot{c}_\mu \hat{P}_\perp \ket{\psi_\mu} = -i \sum_\mu c_\mu (h_\mu - h) \hat{P}_\perp \ket{\psi_\mu} \comma
	\label{eqn:App:B:CoefficientsOfPerturbations}
\end{align}
where $\hat{P}_\perp$ is the projector on the subspace perpendicular to $\ket{\psi}$. 
The state $\ket{\psi}$ is stable if all expansion coefficients $c_\mu$ associated to perturbations orthogonal to $\ket{\psi}$ decay to zero. 

Recall that for a non-degenerate spectrum $\{ h_\mu \}$, the pseudo-steady state $\ket{\psi} = \ket{\psi_\alpha}$ is an eigenstate of $\hat{H}_0 - i \hat{M}$ to eigenvalue $h = h_\alpha$.
Therefore, we can rewrite Eq.~\eqref{eqn:App:B:CoefficientsOfPerturbations} to
\begin{align}
	\forall \mu \neq \alpha : \quad \frac{\d c_\mu}{\d t} = -i (h_\mu - h) c_\mu \fullstop
\end{align}
Hence, the state $\ket{\psi}$ is stable if $\Im (h_\mu - h) \leq 0$ holds for all $\mu \neq \alpha$, \emph{i.e.}, if $h$ is the eigenvalue of the spectrum with the largest imaginary part, as shown in Fig.~\ref{fig:Protocol}(c). 
The decay rate of any state $\ket{\psi_\mu}$ towards $\ket{\psi} = \ket{\psi_\alpha}$ is given by $\Gamma_{\mathrm{rel}\,\mu \to \alpha} = - \Im (h_\mu - h) = \bra{\psi_\mu} \hat{M} \ket{\psi_\mu} - \bra{\psi} \hat{M} \ket{\psi}$, which is the imaginary part of the spectral gap between the two eigenstates $\ket{\psi_\mu}$ and $\ket{\psi} = \ket{\psi_\alpha}$. 
Thus, for a non-degenerate spectrum $\{ h_\mu \}$ there is only one stable pseudo-steady state $\ket{\psi}_\mathrm{ps}$ and the relaxation rate towards it is determined by the smallest imaginary gap between the stable pseudo-steady state and the unstable eigenstates of $\hat{H}_0 - i \hat{M}$, as shown in Fig.~\ref{fig:Protocol}(c).

\section{Nonclassical states in a Kerr oscillator}
The state $\hat{\rho}$ of a quantum system can be represented by the Wigner function $W_{\hat{\rho}}(\alpha) = \Tr [ \hat{\rho} \hat{D}(\alpha) \hat{\Pi} \hat{D}^\dagger(\alpha) ]/\pi$, where $\hat{D}(\alpha) = e^{\alpha \hat{a}^\dagger - \alpha^* \hat{a}}$ is the displacement operator and $\hat{\Pi} = e^{i \pi \hat{a}^\dagger \hat{a}}$ is the parity operator \cite{Tilma-PhysRevLett.117.180401}. 
The Wigner function is a quasi-probability distribution in phase space and negative values of $W_{\hat{\rho}}(\alpha)$ indicate a nonclassical state $\hat{\rho}$ \cite{GerryKnight}.
We now show that the pseudo-steady state $\ket{\psi}_\mathrm{ps}$ of a Kerr oscillator can have a negative Wigner function $W_{\ket{\psi}_\mathrm{ps}}(\alpha)$, whereas the steady-state Wigner function $W_{\hat{\rho}_\mathrm{ss}}(\alpha)$ has been proven to be strictly positive \cite{Kheruntsyan-OptCommun.139.157,Kheruntsyan-JourOptB.2.225,Bartolo-PhysRevA.94.033841}. 
As negativity measure, we use the modulus of the minimum of the Wigner function, $N(\hat{\rho}) = \abs{\min_\alpha [W_{\hat{\rho}}(\alpha)]}$, which is non-zero if $W_{\hat{\rho}}(\alpha)$ takes negative values and zero otherwise.

\subsection{Semiclassical drive}
\begin{figure}
	\centering
	\includegraphics[width=0.48\textwidth]{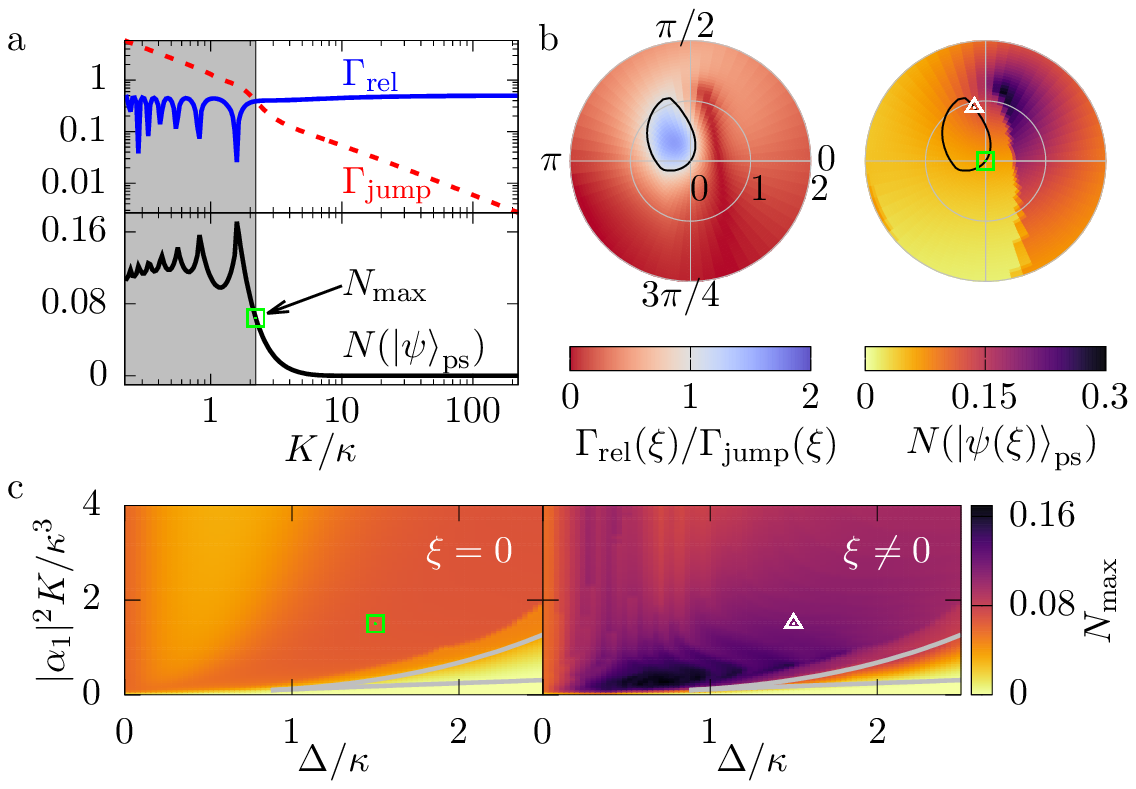}
	\caption{
		(a) Relaxation rate $\Gamma_\mathrm{rel}$ to the  pseudo-steady state $\ket{\psi}_\mathrm{ps}$ (solid blue), jump rate $\Gamma_\mathrm{jump}$ (dotted red), and negativity of the Wigner function (solid black) for a Kerr oscillator subject to a semiclassical drive for fixed dimensionless detuning $\Delta/\kappa = \ValueFigIIdelta$ and rescaled drive power $\abs{\alpha_1}^2 K/\kappa^3 = \ValueFigIIP$. 
		In the area highlighted in gray, the quantum trajectory is dominated by stochastic quantum jumps, $\Gamma_\mathrm{jump} \geq \Gamma_\mathrm{rel}$, and $\ket{\psi}_\mathrm{ps}$ cannot be prepared.
		The open green rectangle indicates the maximum observable negativity $N_\mathrm{max}$ and the parameters of Fig.~\ref{fig:Protocol}. 
		(b) Adding a local oscillator signal $\sqrt{\kappa} \xi$ allows one to unravel different pseudo-steady states.
		The ratio $\Gamma_\mathrm{rel}(\xi)/\Gamma_\mathrm{jump}(\xi)$ (left plot) and the negativity $N(\ket{\psi(\xi)}_\mathrm{ps})$ (right plot) now depend on the complex signal strength $\xi$. 
		All states within the black curve indicating $\Gamma_\mathrm{rel}(\xi)/\Gamma_\mathrm{jump}(\xi) = 1$ can be prepared in a heralded way. 
		The value of $\xi$ indicated by an open white triangle maximizes $N(\ket{\psi (\xi)}_\mathrm{ps})$ under this restriction.
		(c) Maximum observable negativity $N_\mathrm{max}$ as a function of dimensionless detuning and rescaled drive power without (left panel) and with (right panel) an optimization of the local oscillator signal $\xi$. 
		In the triangle enclosed by the gray lines, two semiclassical steady-state solutions $\erw{\hat{a}}$ exist.		
	}
	\label{fig:Semiclassical}
\end{figure}
We consider a semiclassical drive, $\alpha_1 \geq 0$, and set $\alpha_2 = 0$, such that the steady-state solution is characterized by the detuning $\Delta/\kappa$, the rescaled drive power $\abs{\alpha_1}^2 K/\kappa^3$, and the ratio $K/\kappa$ \cite{Meaney-EPJQT.1.7}.
For fixed values of the first two quantities and $K \gg \kappa$, the pseudo-steady state $\ket{\psi}_\mathrm{ps}$ is positive, as shown in Fig.~\ref{fig:Semiclassical}(a). 
This is due to the fact that the steady state of a Kerr oscillator is strictly positive. 
If the relaxation rate dominates, $\Gamma_\mathrm{rel} \gg \Gamma_\mathrm{jump}$, the system is almost always in the pseudo-steady state and, therefore, $\ket{\psi}_\mathrm{ps}$ must be identical to $\hat{\rho}_\mathrm{ss}$ to ensure that an ensemble average over many trajectories reproduces the steady state.
However, if relaxation rate and jump rate are comparable, $\Gamma_\mathrm{rel} \gtrsim \Gamma_\mathrm{jump}$, the pseudo-steady state differs from $\hat{\rho}_\mathrm{ss}$ and can be nonclassical, as shown in Fig.~\ref{fig:Semiclassical}(a).
Quantum jumps let $\ket{\psi}$ explore many different states that compensate the nonclassicality of $\ket{\psi}_\mathrm{ps}$ and average out to a positive steady state.
Finally, for $K \ll \kappa$ the quantum trajectory is dominated by stochastic quantum jump events.
Then, $\ket{\psi}$ can no longer relax to $\ket{\psi}_\mathrm{ps}$ because the intervals between two quantum jumps are much shorter than the relaxation time, $\Gamma_\mathrm{jump} \gg \Gamma_\mathrm{rel}$. 
Considering this, we define the \emph{maximum observable negativity} $N_\mathrm{max}$ as the maximum of $N(\ket{\psi}_\mathrm{ps})$ in the regime $\Gamma_\mathrm{rel} \geq \Gamma_\mathrm{jump}$.
The left panel of Fig.~\ref{fig:Semiclassical}(c) displays $N_\mathrm{max}$ as a function of the dimensionless detuning and the rescaled drive power. 
Usually, the negativity $N(\ket{\psi}_\mathrm{ps})$ decreases monotonically as a function of $K/\kappa$, such that the maximum observable negativity $N_\mathrm{max}$ is achieved for $\Gamma_\mathrm{rel} = \Gamma_\mathrm{jump}$. 
However, in the regime where two stable semiclassical solutions exist, enclosed by the gray lines in Fig.~\ref{fig:Semiclassical}(c), the largest negativity is observed for $\Gamma_\mathrm{rel} > \Gamma_\mathrm{jump}$.

\subsection{Unraveling different states}
The unraveling of the quantum master equation~\eqref{eqn:QME} is not unique \cite{WisemanMilburn}.
Thus, the operator $\super{H}$ is not unique and many different pseudo-stationary states $\ket{\psi}_\mathrm{ps}$ can be stabilized to a given steady-state solution $\hat{\rho}_\mathrm{ss}$. 
To illustrate this point, we consider the homodyne detection setup shown in Fig.~\ref{fig:System}(b). 
A beamsplitter is placed between the system and the photon detector, such that the signal $\sqrt{\kappa} \xi$ of a local oscillator is added to the system's output and the jump probability is modified, $\Exp(\d N') = \kappa \bra{\psi} (\hat{a}^\dagger + \xi^*)(\hat{a} + \xi) \ket{\psi} \d t$. 
This corresponds to a photon-counting measurement in a displaced frame $\ket{\chi} = \hat{D}(\xi) \ket{\psi}$ with a modified Hamiltonian $\hat{H}_0'(\xi) = \hat{D}(\xi) \hat{H}_0 \hat{D}^\dagger(\xi) - i \kappa (\xi^* \hat{a} - \xi \hat{a}^\dagger)/2$.
Thus, the results derived in Sec.~\ref{sec:PSS} can be carried over straightforwardly. 

The local oscillator signal $\sqrt{\kappa} \xi$ now allows us to modify the ratio $\Gamma_\mathrm{rel}(\xi)/\Gamma_\mathrm{jump}(\xi)$ and the pseudo-steady state $\ket{\psi(\xi)}_\mathrm{ps}$, as shown in Fig.~\ref{fig:Semiclassical}(b).
In contrast to the standard homodyne detection limit $\abs{\xi} \gg \erw{\hat{a}}$, where the local oscillator signal dominates and the quantum trajectory is a continuous Wiener process \cite{WisemanMilburn}, we consider the opposite limit  $\abs{\xi} \lesssim \erw{\hat{a}}$, such that the detection of photons is still a Poissonian quantum jump process.
Moreover, a state $\ket{\psi(\xi)}_\mathrm{ps}$ can only be prepared if $\Gamma_\mathrm{rel} \leq \Gamma_\mathrm{jump}$ holds, which restricts $\xi$ to the area inside the black curve in Fig.~\ref{fig:Semiclassical}(b). 
Nevertheless, an optimization of the local oscillator signal $\xi$ under these constraints significantly increases the maximum observable negativity $N_\mathrm{max}$ over the case of $\xi = 0$, as shown in the right panel of Fig.~\ref{fig:Semiclassical}(c).

\subsection{Parametric drive}
\begin{figure}
	\centering
	\includegraphics[width=.48\textwidth]{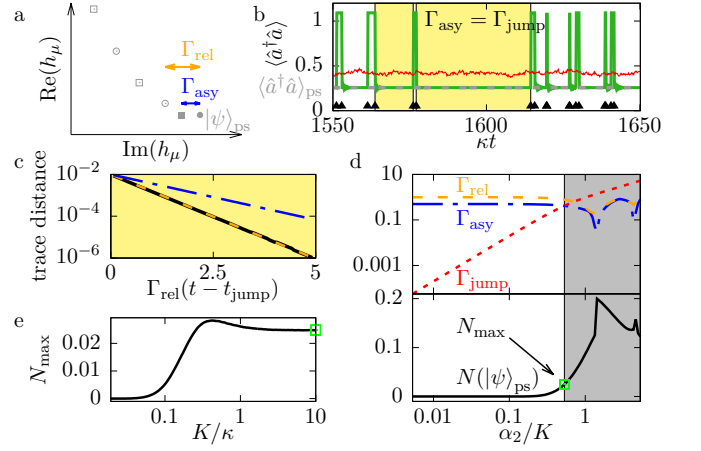}
	\caption{
		(a) Spectrum of even-parity (circles) and odd-parity (squares) stable (solid markers) and unstable (open markers) eigenstates of the non-Hermitian Hamiltonian that defines the relaxation dynamics for a Kerr oscillator subject to a resonant parametric drive.
		The imaginary part of the gap between the two stable states determines their jump-rate asymmetry $\Gamma_\mathrm{asy}$.
		(b) Photon-number $\erw{\hat{a}^\dagger \hat{a}}$ in the steady-state regime. 
		An average over $\ValueFigIIINtraj$ trajectories reproduces the steady-state result (thin red), which determines the jump rate $\Gamma_\mathrm{jump}$. 
		Each quantum trajectory (solid green line) jumps between the stable states of opposite parity (jump times indicated by black triangles).
		If $\Gamma_\mathrm{asy} \gtrsim \Gamma_\mathrm{jump}$ holds, one can prepare the stable even-parity eigenstate $\ket{\psi}_\mathrm{ps}$ in a heralded way.
		(c) After a quantum jump event, the trace distance between $\ket{\psi(t)}$ and $\ket{\psi}_\mathrm{ps}$ (solid black) decays exponentially. 
		Since parity is conserved, the relaxation happens at a rate $\Gamma_\mathrm{rel}$ (dashed orange), which is the imaginary part of the second spectral gap. 
		For comparison, the dash-dotted blue line indicates a decay at the rate $\Gamma_\mathrm{asy}$ corresponding to the first spectral gap.		
		(d) Relaxation rate $\Gamma_\mathrm{rel}$ (dashed orange), jump-rate asymmetry $\Gamma_\mathrm{asy}$ (dash-dotted blue), total jump rate $\Gamma_\mathrm{jump}$ (dashed red), and Wigner-function negativity (solid black) as a function of the drive strength. 
		In the gray area, the time evolution is dominated by stochastic quantum jumps, $\Gamma_\mathrm{jump} \geq \Gamma_\mathrm{asy}$, and $\ket{\psi}_\mathrm{ps}$ cannot be prepared. 
		(e) Maximum observable negativity as a function of the dimensionless Kerr nonlinearity $K/\kappa$. 
		Parameters: $\Delta/\kappa = \ValueFigIIIDelta$, $K/\kappa = \ValueFigIIIK$, $\alpha_1/\kappa = \ValueFigIIIaI$, $\alpha_2/\kappa = \ValueFigIIIaII$, and $\xi = \ValueFigIIIxi$. 
	}
	\label{fig:Parametric}
\end{figure}
Our protocol can be used to stabilize a Schr\"odinger kitten state in a Kerr oscillator without the need for feedback \cite{Minganti-SciRep.6.26987}:
We consider a resonant parametric drive, \emph{i.e.}, $\Delta = 0$, $\alpha_1 = 0$, and $\alpha_2 \geq 0$, such that the non-Hermitian Hamiltonian $\hat{H}_0 - i \hat{M}$ commutes with the parity operator $\hat{\Pi}$ and the spectrum consists of two subspaces of eigenstates having different parity, $\{ h_\mu^\pm \}$.
The operator $\super{H}$ does not mix these subspaces, therefore, both the even and the odd-parity eigenstate $\ket{\psi_{\mu_0}^\pm}$ with largest imaginary part of the eigenvalue $h_{\mu_0}^\pm$ are stable, as shown in Fig.~\ref{fig:Parametric}(a), and their relaxation rates are determined by the imaginary parts of the spectral gaps to the unstable eigenstates of the corresponding parity. 

While we redefined here the relaxation rate $\Gamma_{\mathrm{rel}}$ to take into account parity conservation, the relevant quantity to be compared to $\Gamma_\mathrm{jump}$ in the heralding protocol is still the \emph{first} spectral gap, $\Gamma_\mathrm{asy} = \Im (h_{\mu_0}^+ - h_{\mu_0}^-)$:
Photon detection events change the parity of $\ket{\psi}$ and approximately map the stable states $\ket{\psi_{\mu_0}^\pm}$ to one another, such that the quantum trajectories jump between the two states, as shown in Fig.~\ref{fig:Parametric}(b). 
The rate $\Gamma_\mathrm{asy}$ measures the asymmetry in the jump rates of $\ket{\psi_{\mu_0}^\pm}$, which reflects their different photon-number expectation values. 
If $\Gamma_\mathrm{asy} \geq \Gamma_\mathrm{jump}$ holds, the states can be discriminated in the photon detection signal and the longer-lived state $\ket{\psi_{\mu_0}^+}$ can be prepared in a heralded way, $\ket{\psi}_\mathrm{ps} = \ket{\psi_{\mu_0}^+}$.
The relaxation rate $\Gamma_\mathrm{rel}$ towards $\ket{\psi}_\mathrm{ps}$ is given by the \emph{second} spectral gap and determines the relaxation $\hat{a} \ket{\psi_{\mu_0}^-} \to \ket{\psi_{\mu_0}^+}$, as shown in Fig.~\ref{fig:Parametric}(c). 
Since $\Gamma_\mathrm{rel} > \Gamma_\mathrm{asy}$ holds, the relaxation to the target state within the heralding interval is guaranteed. 
Similar to the case of a semiclassical drive, Figs.~\ref{fig:Parametric}(d) and~(e) show that $\ket{\psi}_\mathrm{ps}$ can have a negative Wigner function if $\Gamma_\mathrm{asy} \approx \Gamma_\mathrm{jump}$ and $K \gtrsim \kappa$ hold, but $N(\ket{\psi}_\mathrm{ps})$ is zero in the limit $\Gamma_\mathrm{asy} \gg \Gamma_\mathrm{jump}$ because $\ket{\psi}_\mathrm{ps}$ converges to the positive steady state $\hat{\rho}_\mathrm{ss}$.
Note that the convergence $\hat{\rho}_\mathrm{ps} \to \hat{\rho}_\mathrm{ss}$ if $\Gamma_\mathrm{rel},\Gamma_\mathrm{asy} \gg \Gamma_\mathrm{jump}$ is specific to the Kerr oscillator studied here.

Importantly, in the limit $K \gg \kappa$ the states $\ket{\psi_{\mu_0}^\pm}$ converge to the even and odd Schr\"odinger cat states $\ket{\mathcal{C}_\pm} = (\ket{\alpha} \pm \ket{-\alpha})/[2 (1 \pm e^{-2 \abs{\alpha}^2})]^{1/2}$ \cite{GerryKnight}, where $\alpha = i \sqrt{\alpha_2/K}$.
In this regime, the steady-state solution $\hat{\rho}_\mathrm{ss}$ is a statistical mixture of the two indistinguishable cat states $\ket{\mathcal{C}_\pm}$. 
The small correction $-i \hat{M} \propto \kappa$ due to the photon detection breaks this symmetry and allows us to stabilize the  even-parity Schr\"odinger kitten state $\ket{\mathcal{C}_+} = \ket{\psi}_\mathrm{ps}$ without feedback.

\section{Finite temperature and imperfect photon detection}
\label{sec:FINITE}
In an experiment, the environment will be at finite temperature and it may emit photons into the open quantum system. 
Moreover, current photon detectors have detection efficiencies of less than $100\,\%$ such that photons emitted by the open quantum system may escape undetectedly. 
Both effects give rise to unmonitored dissipative processes that require us to go beyond the stochastic Schr\"odinger equation~\eqref{eqn:SSE}, \emph{i.e.}, we must describe the system by a stochastic master equation for a density matrix $\hat{\rho}$. 
In Appendix~\ref{sec:app:PseudoStationaryStateAndRelaxationRate}, we discuss this generalization of our findings to finite temperature $n_\mathrm{th} > 0$, imperfect detection $0 \leq \eta < 1$, and additional dissipative channels in $\super{L}_0$. 
In essence, the unobserved dissipative processes mix different eigenstates of $\hat{H}_0 - i \hat{M}$, the pseudo-steady state becomes a mixed state $\hat{\rho}_\mathrm{ps}$, and negativities in the Wigner function get averaged out depending on the statistical mixture of eigenstates described by $\hat{\rho}_\mathrm{ps}$. 

\begin{figure}
	\centering
	\includegraphics[width=0.48\textwidth]{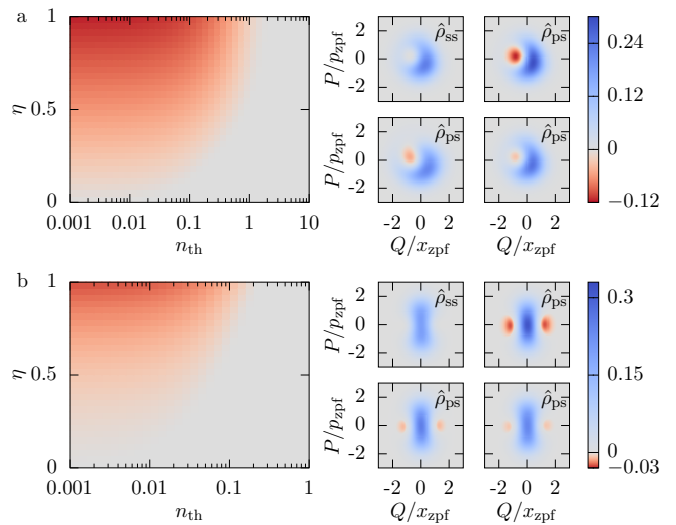}
	\caption{
		(a) Impact of finite temperature or imperfect detection on the pseudo-steady state of a Kerr oscillator subject to a semiclassical drive. 
		The main plot shows the minimum of the Wigner function $\min_\alpha[ W_{\hat{\rho}_\mathrm{ps}}(\alpha)] = - N(\hat{\rho}_\mathrm{ps})$ as a function of the thermal photon number $n_\mathrm{th}$ and the detection efficiency $\eta$. 
		The smaller plots show the Wigner function $W_{\hat{\rho}}(\alpha)$ of selected states. 
		The origin has been shifted to the steady-state expectation value $\erw{\hat{a}}_\mathrm{ss}$. 
		Top row: Wigner function of steady-state $\hat{\rho}_\mathrm{ss}$ and pseudo-steady state $\hat{\rho}_\mathrm{ps}$ for $n_\mathrm{th} = \ValueFigIVSCnthI$ and $\eta = \ValueFigIVSCetaI$. 
		Bottom row: Wigner function of pseudo-steady state $\hat{\rho}_\mathrm{ps}$ for $n_\mathrm{th} = \ValueFigIVSCnthII$ and $\eta = \ValueFigIVSCetaII$ (left) and $n_\mathrm{th} = \ValueFigIVSCnthIII$ and $\eta = \ValueFigIVSCetaIII$ (right). 
		Parameters are $\Delta/\kappa = \ValueFigIVSCdelta$, $\abs{\alpha_1}^2 K/\kappa^3 = \ValueFigIVSCP$, $\alpha_2/\kappa = \ValueFigIVSCaII$, $K/\kappa = \ValueFigIVSCK$, $\xi/\sqrt{\kappa} = \ValueFigIVSCxiabs \times \exp(\ValueFigIVSCxiarg i)$. 
		(b) Same plots for a Kerr oscillator subject to a parametric drive. 
		Top row: Wigner function of steady-state $\hat{\rho}_\mathrm{ss}$ and pseudo-steady state $\hat{\rho}_\mathrm{ps}$ for $n_\mathrm{th} = \ValueFigIVPAnthI$ and $\eta = \ValueFigIVPAetaI$. 
		Bottom row: Wigner function of pseudo-steady state for $n_\mathrm{th} = \ValueFigIVPAnthII$ and $\eta = \ValueFigIVPAetaII$ (left) and $n_\mathrm{th} = \ValueFigIVPAnthIII$ and $\eta = \ValueFigIVPAetaIII$ (right). 
		Parameters are $\Delta/\kappa = \ValueFigIVPADelta$, $\alpha_1/\kappa = \ValueFigIVPAaI$, $\alpha_2/\kappa = \ValueFigIVPAaII$, $K/\kappa = \ValueFigIVPAK$, $\xi= 0$.
	}
	\label{fig:Finite}
\end{figure}

In Fig.~\ref{fig:Finite}, the minimum of the Wigner function, $\min_\alpha [ W_{\hat{\rho}_\mathrm{ps}}(\alpha)] = - N(\hat{\rho}_\mathrm{ps})$, is shown for finite temperature or imperfect photon detection.
Note that imperfect photon detection both includes a loss of photons on the way to the detector and a detection efficiency less than unity at the detector itself.
Thermal effects average out the negativity at a thermal photon number of about $n_\mathrm{th} \approx 0.1$.
Hence, negative Wigner functions can be observed in the optical frequency range, but pre-cooling or cryogenic environments are necessary for microwave-frequency setups.

Importantly, Fig.~\ref{fig:Finite} demonstrates that imperfect photon detection is not a major challenge.
Even for a relatively low detection efficiency of $\eta \approx 0.25$ for a semiclassical drive and $\eta \approx 0.5$ for a parametric drive, negativities in the Wigner function are still present. 
Thus, current photon detection efficiencies in the optical and infrared range of above $88\%$ are promising to resolve nonclassical states \cite{Takeuchi-ApplPhysLett.74.1063,Fukuda-OptExpress.19.870}. 
The single-photon detection efficiency in the microwave regime is still lower \cite{Munro-PhysRevA.71.033819}, but recently values exceeding $70\%$ have been reached \cite{Besse-PhysRevX.8.021003,Kono-nphys.14.546}.

\section{Experimental implementation}
\label{sec:EXP}
Our results show that quantum oscillators with Kerr nonlinearities of the order of the decay rate $\kappa$ are sufficient to observe negative pseudo-steady state Wigner functions. 
Such nonlinear resonators can be realized in a variety of platforms, \emph{e.g.}, superconducting circuits \cite{Rimberg-NJP.16.055008,Heikkila-PhysRevLett.112.203603} and trapped ions \cite{Babikov-PhysRevA.77.012338,Home-NJP.13.073026}. 
Potentially, even hybrid optomechanical systems could reach the required nonlinearities \cite{Jacobs-PhysRevLett.103.067201,Zhang-PhysRevB.92.115407,Rimberg-NJP.16.055008,YiwenChu}.
To ensure $\Gamma_\mathrm{jump} \lesssim \Gamma_\mathrm{rel}$, the steady-state photon number needs to be small, $\erw{\hat{a}^\dagger \hat{a}}_\mathrm{ss} \lesssim 1$. 

A first step towards an experimental realization of our proposal is to demonstrate the nonclassicality of the pseudo-steady state in a Wigner function tomography.
This can be achieved with minimal complexity using the existing setup shown in Fig.~\ref{fig:System}(b) and the protocols described in Refs.~\onlinecite{PhysRevA.53.4528,PhysRevLett.76.4344}. 
These protocols are based on the fact that the value of the Wigner function $W(0)$ at the origin of phase space can be obtained by simple photon detection. 
A displacement of the mode prior to detection allows one to measure the Wigner function at different positions in phase space. 
A possible experiment will consist of repeated runs of data collection, each one measuring one pixel $W(\alpha)$ of the Wigner function. 
A run starts with a state preparation step as described in Sec.~\ref{sec:PRT}. 
When the generation of a nonclassical state is heralded, the tomography step begins and the local oscillator signal is suddenly changed to displace the state and measure the Wigner function at the coordinate $\alpha$.

Other ways to perform a Wigner function tomography have been demonstrated in experiments with superconducting circuits or trapped ions. 
There, one measures the interaction of an (artificial) atom with the nonclassical quantum state to reconstruct the Wigner function \cite{PhysRevLett.91.010401,nature.459.546,PhysRevLett.110.100404}. 

Having verified the nonclassicality of the pseudo-steady state, the next experimental step is to extract and use it. 
Hybrid optomechanical systems are promising candidates for this task, because they allow one to perform on-demand state-swap operations between their modes \cite{RevModPhys.86.1391}. 
An experimental protocol could consist of a state preparation step in an optical mode of the system, followed by a state swap to another mode if the presence of a nonclassical state is heralded. 
The properties of the target mode of the state swap are tailored to the task one wishes to perform with the nonclassical state.

\section{Conclusion}
We have shown that continuous photon detection can stabilize nonclassical pseudo-steady states in a driven and damped Kerr nonlinear oscillator, whose steady-state Wigner function is known to be strictly positive. 
The required nonlinearities and photon detection efficiencies are feasible with current technology.
We have applied this protocol to a Kerr parametric oscillator to prepare Schr\"odinger kitten states.
Making use of the jump-rate asymmetry between the states of different parity, we demonstrated that observation is sufficient to stabilize such nonclassical states, even in the absence of feedback.
Finally, viewed from a different angle, the proposed scheme is a heralding protocol to stabilize quantum states in open systems.

\begin{acknowledgments}
We thank P.\ Treutlein and K.\ Hammerer for fruitful discussions. 
This work was financially supported by the Swiss National Science Foundation (SNSF) and the NCCR Quantum Science and Technology. 
The numerical simulations have been implemented using the qutip package \cite{qutip}. 
Parts of the calculations were performed at the sciCORE scientific computing core facility at University of Basel \cite{scicore}.
\end{acknowledgments}

\appendix

\section{Pseudo-steady state of a stochastic quantum master equation}
\label{sec:app:PseudoStationaryStateAndRelaxationRate}
In this Appendix, we consider the general case of an unraveling of the quantum master equation~\eqref{eqn:QME} where $\super{L}_0$ is any completely positive and trace preserving linear superoperator such that Eq.~\eqref{eqn:QME} has a steady-state solution $\hat{\rho}_\mathrm{ss}$.
We assume that the output mode $\hat{a}$ is displaced by a local oscillator signal of strength $\sqrt{\kappa (n_\mathrm{th} + 1) \eta} \xi$ before photon detection, as sketched in Fig.~\ref{fig:System}(b). 
Note that the case $\xi = 0$ reproduces the conventional photon-detection scenario. 
Under these more general conditions, the corresponding stochastic quantum master equation is given by \cite{WisemanMilburn}
\begin{align}
		\d \hat{\rho} 
			&= \overline{\super{L}} \hat{\rho} \, \d t 
			+ \left[ \frac{(\hat{a} + \xi) \hat{\rho} (\hat{a}^\dagger + \xi^*)}{\Tr [ (\hat{a}^\dagger + \xi^*)(\hat{a} + \xi) \hat{\rho}]} - \hat{\rho} \right] \d N \comma
		\label{eqn:App:A:SME} \\
		\overline{\super{L}} \hat{\rho} 
			&= ( \super{L} + \super{N} ) \hat{\rho} - \Tr (\super{N} \hat{\rho}) \hat{\rho} \comma
\end{align}
where we introduced the abbreviations
\begin{align}
	\super{L} \hat{\rho} 
		&= \super{L}_0 \hat{\rho} - i \komm{\kappa (n_\mathrm{th} + 1) \eta \frac{i}{2} (\xi \hat{a}^\dagger - \xi^* \hat{a})}{\hat{\rho}} \nonumber \\
		&+ \kappa (n_\mathrm{th} + 1) (1 - \eta) \D{\hat{a}} \hat{\rho} + \kappa n_\mathrm{th} \D{\hat{a}^\dagger} \hat{\rho} \comma 
		\label{eqn:App:A:L} \\
	\super{N} \hat{\rho} 
		&= - \frac{\kappa}{2} (n_\mathrm{th} + 1) \eta \akomm{(\hat{a}^\dagger + \xi^*)(\hat{a} + \xi)}{\hat{\rho}} \fullstop	
\end{align}
The Poissonian increment $\d N = \d N^2$ has the ensemble-averaged expectation value $\Exp(\d N) = - \Tr(\super{N} \hat{\rho}) \d t$. 
The superoperator $\super{N} \hat{\rho}$ is the counterpart of the non-Hermitian operator $-i \hat{M}$ in Eq.~\eqref{eqn:HNL}, \emph{i.e.}, it describes the modification of the dynamics if no photons are detected and causes a decay of the trace of $\hat{\rho}$. 
To preserve the normalization, we include the nonlinear term $- \Tr(\super{N} \hat{\rho}) \hat{\rho}$ into $\overline{\super{L}}\hat{\rho}$. 

In the following we require that the quantum master equation~\eqref{eqn:QME} has a steady-state solution $\hat{\rho}_\mathrm{ss}$ and that the superoperator $\super{L} + \super{N}$ has a set of left and right eigenvectors
\begin{align}
	(\super{L} + \super{N}) \hat{\rho}_\mu &= \lambda_\mu \hat{\rho}_\mu 
	\label{eqn:App:A:BasisMatrices}\\
	(\super{L} + \super{N})^\dagger \check{\rho}_\mu &= \lambda_\mu^* \check{\rho}_\mu 
\end{align}
that can suitably be normalized to form a complete orthonormal basis with respect to the Hilbert-Schmidt scalar product, $\left( \check{\rho}_\nu, \hat{\rho}_\mu \right) = \Tr \left( \check{\rho}_\nu^\dagger \hat{\rho}_\mu \right) = \delta_{\nu,\mu}$.
This assumption is valid for all systems that do not have exceptional points \cite{Moiseyev}.

In the limit $\super{L}_0 \hat{\rho} \to -i \komm{\hat{H}}{\hat{\rho}}$, $n_\mathrm{th} \to 0$, and $\eta \to 1$, the stochastic Schr\"odinger equation~\eqref{eqn:SSE} and the stochastic quantum master equation~\eqref{eqn:App:A:SME} can be mapped onto one another. 
The right eigenstates $\ket{\psi_j}$ of $\hat{H} - i \hat{M}$, cf.\ Eq.~\eqref{eqn:App:B:BasisKets}, can be used to construct the right eigenstates $\hat{\rho}_\mu = \hat{\rho}_{i,j} = \ket{\psi_i}\bra{\psi_j}$ of $\super{L} + \super{N}$, cf.\ Eq.~\eqref{eqn:App:A:BasisMatrices}, and the corresponding eigenvalues fulfill $\lambda_\mu = \lambda_{i,j} = -i (h_i - h_j^*)$.   
For finite temperature $n_\mathrm{th} > 0$, imperfect detection efficiency $0 \leq \eta < 1$, or additional dissipation channels in $\super{L}_0$, this relation breaks down because the additional Lindblad dissipators in Eq.~\eqref{eqn:App:A:L} mix different basis states $\hat{\rho}_{i,j}$. 
Note that non-Hermitian states $\hat{\rho}_{i, j \neq i}$ are never mixed with Hermitian states $\hat{\rho}_{i,i}$ because $\overline{\super{L}} \hat{\rho}$ must preserve the Hermiticity of $\hat{\rho}$.
Physically, these processes correspond to unmonitored dissipative interactions such that the system state can no longer be described by a pure state $\ket{\psi}$. 
Instead, different states $\hat{\rho}_{i,i}$, each of them possibly having a negative Wigner function, are mixed and their negativity is ultimately averaged out to a non-negative pseudo-steady-state Wigner function in the limit $\eta \to 0$.

The pseudo-steady state of Eq.~\eqref{eqn:App:A:SME} is a density matrix $\hat{\rho}$ that is Hermitian, positive semidefinite, normalized to unit trace, and that satisfies $\overline{\super{L}} \hat{\rho} = 0$.
In analogy to the treatment in the main text, we decompose $\hat{\rho}$ with respect to the basis of eigenstates of $\super{L} + \super{N}$, $\hat{\rho} = \sum_\mu c_\mu \hat{\rho}_\mu$, and obtain the following conditions for the expansion coefficients:
\begin{align}
	\forall \mu : \quad c_\mu \left[ \lambda_\mu - \sum_\beta c_\beta \lambda_\beta \Tr(\hat{\rho}_\beta) \right] = 0 \fullstop
	\label{eqn:App:A:Coefficients}
\end{align}
For a non-degenerate eigenvalue $\lambda_\nu$, all but the coefficient $c_\nu$ of the corresponding eigenstate $\hat{\rho}_\nu$ must be zero. 
Thus, each eigenstate $\hat{\rho}_\nu$ to a non-degenerate eigenvalue $\lambda_\nu$ is a valid solution provided that it is Hermitian, positive semidefinite, and has a non-zero trace such that it can be normalized by $c_\nu = 1/\Tr(\hat{\rho}_\nu)$.
For a degenerate eigenvalue $\lambda = \lambda_{\nu_1} = \dots = \lambda_{\nu_N}$, only the coefficients $c_{\nu_i}$ of eigenstates $\hat{\rho}_{\nu_i}$ belonging to the degenerate subspace $\{ \lambda_{\nu_1}, \dots, \lambda_{\nu_N}\}$ are non-zero. 
Any mixture $\hat{\rho} = \sum_{i=1}^N c_{\nu_i} \hat{\rho}_{\nu_i}$ of these eigenstates is a valid solution provided that it is Hermitian, positive semidefinite, and normalized to unit trace, $\sum_{i=1}^N c_{\nu_i} \Tr( \hat{\rho}_{\nu_i}) = 1$.
It can be shown that convex combinations of eigenstates $\ket{\psi_j}$ with real eigenvalues are a solution to $\overline{\super{L}} \hat{\rho} = 0$ \cite{Graefe}.

Since $\overline{\super{L}} \hat{\rho}$ is a nonlinear superoperator, some of the solutions to $\overline{\super{L}} \hat{\rho} = 0$ determined above may be unstable against perturbations. 
To analyze the stability of a solution $\hat{\rho}$ to eigenvalue $\lambda$, we make the ansatz
\begin{align}
	\hat{\chi} = (\hat{\rho} + \varepsilon \hat{\sigma}) [ 1 - \varepsilon \Tr(\hat{\sigma}) ] \comma
\end{align}
where $\varepsilon \ll 1$ is a small parameter and $\hat{\sigma}$ is a Hermitian and positive-semidefinite density matrix that is orthogonal to $\hat{\rho}$. 
Note that $\hat{\chi}$ is normalized to leading order in $\varepsilon$. 
We expand $\dot{\hat{\chi}} = \overline{\super{L}} \hat{\chi}$ in terms of $\varepsilon$ and decompose $\hat{\sigma} = \sum_\mu c_\mu \hat{\rho}_\mu$ with respect to the basis of eigenstates of $\super{L} + \super{N}$, which yields
\begin{align}
	\sum_\mu \dot{c}_\mu \super{P}_\perp \hat{\rho}_\mu = \sum_\mu c_\mu (\lambda_\mu - \lambda) \super{P}_\perp \hat{\rho}_\mu \comma
	\label{eqn:App:A:CoefficientsOfPerturbations}
\end{align}
where $\super{P}_\perp$ is the projector on the subspace perpendicular to $\hat{\rho}$. 
The state $\hat{\rho}$ is stable if all expansion coefficients $c_\mu$ of perturbations orthogonal to $\hat{\rho}$ decay to zero.

For a non-degenerate spectrum $\{ \lambda_\mu \}$, $\hat{\rho} = \hat{\rho}_\alpha$ is an eigenstate of $\super{L} + \super{N}$ to eigenvalue $\lambda = \lambda_\alpha$ and we can rewrite Eq.~\eqref{eqn:App:A:CoefficientsOfPerturbations} to
\begin{align}
	\forall \mu \neq \alpha : \quad \frac{\d c_\mu}{\d t} = (\lambda_\mu - \lambda) c_\mu \fullstop
\end{align}
Hence, the state $\hat{\rho} = \hat{\rho}_\alpha$ is stable if $\Re(\lambda_\mu - \lambda) \leq 0$ holds for all $\mu \neq \alpha$, \emph{i.e.}, if $\lambda$ is the eigenvalue of the spectrum with the largest real part.

\end{document}